\documentclass[12pt]{article}
\usepackage{graphicx}
\usepackage{mathtools}
\usepackage{gensymb}
\usepackage{lineno}
\usepackage{textcomp}
\usepackage{lineno}
\newcommand\pubnumber{DPF2015-298}
\newcommand\pubdate{\today}

\def\Title#1{\begin{center} {\Large #1 } \end{center}}
\def\Author#1{\begin{center}{ \sc #1} \end{center}}
\def\Address#1{\begin{center}{ \it #1} \end{center}}

\newcommand\pubblock{\rightline{\begin{tabular}{l} \pubnumber\\
         \pubdate  \end{tabular}}}
\newenvironment{Abstract}{\begin{quotation}  }{\end{quotation}}
\newenvironment{Presented}{\begin{quotation} \begin{center} 
             PRESENTED AT\end{center}\bigskip 
      \begin{center}\begin{large}}{\end{large}\end{center} \end{quotation}}
\def\Acknowledgments{\bigskip  \bigskip \begin{center} \begin{large}
             \bf ACKNOWLEDGMENTS \end{large}\end{center}}
\def\beq{\begin{equation}}
\def\eeq#1{\label{#1}\end{equation}}
\def\eeqn{\end{equation}}

\def\beqa{\begin{eqnarray}}
\def\eeqa#1{\label{#1}\end{eqnarray}}
\def\eeqan{\end{eqnarray}}

\let\bar=\overbar

\def\Dslash{\not{\hbox{\kern-4pt $D$}}}
\def\dslash{\not{\hbox{\kern-2pt $\del$}}}

\def\msb{{\bar{\ssstyle M \kern -1pt S}}}

\begin{document}
\begin{titlepage}
\pubblock

\vfill
\Title{A Deep Observation of Gamma-ray Emission from Cassiopeia A using VERITAS}
\vfill
\Author{ Augusto Ghiotto, for the VERITAS Collaboration}
\Address{Columbia University,
              augusto.ghiotto@columbia.edu}
\vfill
\begin{Abstract}
Supernova remnants (SNRs) have long been considered the leading candidates for the accelerators of cosmic rays within the Galaxy through the process of diffusive shock acceleration. The connection between SNRs and cosmic rays is supported by the detection of high energy (HE; 100 MeV to 100 GeV) and very high energy (VHE; 100 GeV to 100 TeV) gamma rays from young and middle-aged SNRs. However, the interpretation of the gamma-ray observations is not unique. This is because gamma rays can be produced both by electrons through non-thermal Bremsstrahlung and inverse Compton scattering, and by protons through proton-proton collisions and subsequent neutral-pion decay. To disentangle and quantify the contributions of electrons and protons to the gamma-ray flux, it is necessary to measure precisely the spectra and morphology of SNRs over a broad range of gamma-ray energies. Cassiopeia A (Cas A) is one such young SNR ($\sim 350$ years) which is bright in radio and X-rays. It has been detected as a bright point source in HE gamma rays by Fermi-LAT and in VHE gamma rays by HEGRA, MAGIC and VERITAS. Cas A has been observed with VERITAS for more than 60 hours, tripling the published exposure. The observations span 2007-2013, and half of the data were taken at large zenith angles to boost the effective area above a few TeV. We will present the detailed spectral and morphological results from the complete dataset.
\end{Abstract}
\vfill
\begin{Presented}
DPF 2015\\
The Meeting of the American Physical Society\\
Division of Particles and Fields\\
Ann Arbor, Michigan, August 4--8, 2015\\
\end{Presented}
\vfill
\end{titlepage}
\def\thefootnote{\fnsymbol{footnote}}
\setcounter{footnote}{0}

\section{Introduction}

   Cassiopeia A (Cas A) is a young supernova remnant ( $\sim 350$ years~\cite{ashworth}) at distance of $\sim 3.4$ kpc~\cite{reed} from Earth. It is currently thought, based on light echo observations, that the remnant was created by a Type IIb supernova of 15-25 solar masses~\cite{krause}. This type of supernova occurs when a red supergiant that has already lost its hydrogen envelope undergoes core collapse. 
	Objects such as Cas A have been considered among the best candidates to explain the acceleration of cosmic rays, whose sources cannot be resolved due to diffusion by magnetic fields. Gamma rays, on the other hand, can be traced back to their point of origin, offering an avenue to understanding particle acceleration in supernova remnants (SNRs) and other astrophysical objects.
	Gamma rays in a SNR, however, can be produced either leptonically or hadronically. Leptonic processes include non-thermal bremsstrahlung and inverse-Compton scattering, while hadronic collisions lead to neutral-pion production and decay. Looking for the neutral-pion decay signature in the SNR spectrum can give strong evidence for cosmic ray acceleration in the expanding shell.
	The emission from Cas A has been well studied from radio to gamma rays~\cite{bell, baars}. Cas A is a faint optical source, due mostly to thermal emission in the reverse-ejecta and fast moving knots~\cite{fesen}. In the milimeter wavelength range, the Heinrich Hertz Submillimeter Telescope found a broadening of CO emission lines in the south and west regions of Cas A, indicating interaction between the shock front and nearby molecular clouds. Infrared observations from the Spitzer Space Telescope suggested interaction with molecular clouds in the northern region~\cite{hines, kilpatrick}. 
	Cas A has also been extensively studied in X-rays by XMM-Newton and Chandra (0.1 to 10 keV) and by NuSTAR (3 to 79 keV). A model with two coexisting X-ray emission mechanisms, thermal and non-thermal, is consistent with data. Thermal X-ray emission originates in the reverse-shocked ejecta, rich in highly ionized atoms~\cite{fabian, laming}. Non-thermal - mostly synchrotron - radiation is produced at both the forward and reverse shocks ~\cite{gotthelf, uchiyama}. NuSTAR observations also showed interior knots as sources for non-thermal X-ray emission above 15 keV~\cite{grefenstette}. 
	Very high energy (VHE) gamma rays from Cas A were first detected by HEGRA in 2001~\cite{aharonian}, and later confirmed by MAGIC~\cite{albert} and VERITAS~\cite{acciari}. The spectral index $( \gamma = 2.4 - 2.6)$ and the fluxes at $3\% $ of the Crab Nebula published by these three groups are in good agreement within errors. High energy (HE) gamma rays (MeV to GeV) were first detected by Fermi-LAT in 2010~\cite{abdo}. Subsequent Fermi-LAT data indicated a break in the spectrum around 1.72 GeV, favoring a hadronic emission model over a leptonic one~\cite{yuan, saha}.
	In this study with VERITAS data, we extend the VHE spectrum up to 7 TeV while setting an upper limit in the 10 TeV bin and we look at the centroid location of Cas A. After collecting more than 60 hours of data from 2007 to 2013, we were able to reduce the statistical errors in the spectral index and the centroid below the level of their systematic errors. These updated results have also recently been presented at the 34th ICRC~\cite{kumar}.

\section{The VERITAS Experiment}

The Very Energetic Radiation Imaging Telescope Array System (VERITAS) consists of 4 ground-based telescopes located in southern Arizona ($31\degree $ 40'N, $110\degree $ 57' W, 1.3 km a.s.l.). Each 12-m diameter telescope contains a camera with 499 photomultiplier tubes (PMTs), yielding a $3.5 \degree $ field of view. From 2007 to 2012, there were two major upgrades: 
\begin{itemize}
\item in 2009, a telescope was moved to make the array more symmetric and increase the typical telescope baselines; 
\item 2011-12: installation of FPGA-based camera trigger system and high-efficiency PMTs~\cite{zitzer, kieda}.
\end{itemize}
Currently, a source with a flux level of $1\%$ of the Crab Nebula can be detected in less than 25 h. The angular resolution for gamma rays at 1 TeV is $0.08\degree $ and the sensitivity range spans from 85 GeV to 30 TeV. There are currently 54 sources detected by VERITAS.
	After data are collected, they are analyzed in the following steps:
\begin{itemize}
\item Image is calibrated and cleansed, selecting pixels with Cherenkov light and removing the ones with night sky background~\cite{cogan}; 
\item Hillas parameters are calculated (length, width and size of the image), and used to differentiate showers originated by gamma rays from those originated by cosmic rays~\cite{hillas};
\item The intersection of major axes of the shower images in the camera plane provides a geometric technique to locate the origin of the gamma-ray. 
\end{itemize}

\section{Cas A data}
From 2007 to 2013, more than 60 hours of Cas A data was taken by VERITAS. In this work, we re-analyzed the previously published data~\cite{acciari}, taken at small zenith angles (SZA), and added post-upgrade data taken both at SZA and large zenith angles (LZA). Data were selected undeer dark and clear sky conditions, with all four telescopes functioning and set at $0.5 \degree $ wobble from the source location. LZA data gives a higher effective area for events above 1 TeV in comparison with SZA. Separately, using only post-upgrade SZA data, we looked for the centroid location of the remnant. We summarize our data in Table \ref{table:DataSum}.

   \begin{table}[h!]
\centering
      \caption[]{Summary of Cas A data taken by VERITAS from 2007 to 2013.}
         \label{table:DataSum}
         \begin{tabular}{ccccc}
	\hline
	\hline
             Date   & $ \theta _Z  $ range & Average $ \theta _Z  $ & Live Time & Mean Trigger Rate \\
	      & (deg) & (deg) & (Hours) & (Hz) \\
	\hline      
             09/07 - 11/07 & 27-40 & 34 & 18 & 250 \\
            12/11 - 12/11 & 33-43 & 38 & 2 & 350   \\
            09/12 - 12/12 & 24-39 & 30 & 19 & 400 \\
            09/12 - 12/13 & 40-64 & 56 & 25 & 300   \\
         
         \end{tabular}

   \end{table}

\section{Results and Discussion}

We obtained the spectrum for the entire data set and fitted it with a power-law in the energy range from 300 GeV to 7 TeV, giving a $\chi ^2 $ of 2.22 for 5 degrees of freedom and resulting in a good fit probability of $ 81\% $ (Figure \ref{fig:spectrum} ). The differential energy spectrum for the whole data set is given below:

\begin{equation}
\frac{dN}{dE} = (1.45\pm 0.11) \times 10^{-12} (E/1 TeV)^{-2.75\pm 0.10_{stat} \pm 0.20 _{sys}} cm^{-2}s^{-1}TeV^{-1}
\end{equation}

%%%%%%%%%%%%%%%%%%%%%%%%%%%%%%%%%%%%%%%%%%%%%%%%%%%%%%%%%%%%%%%%%%%%%%%%%
%%
%%   use this format to include an .pdf figure into your paper
%%%
\begin{figure}[h]
\centering
\includegraphics[height=3in]{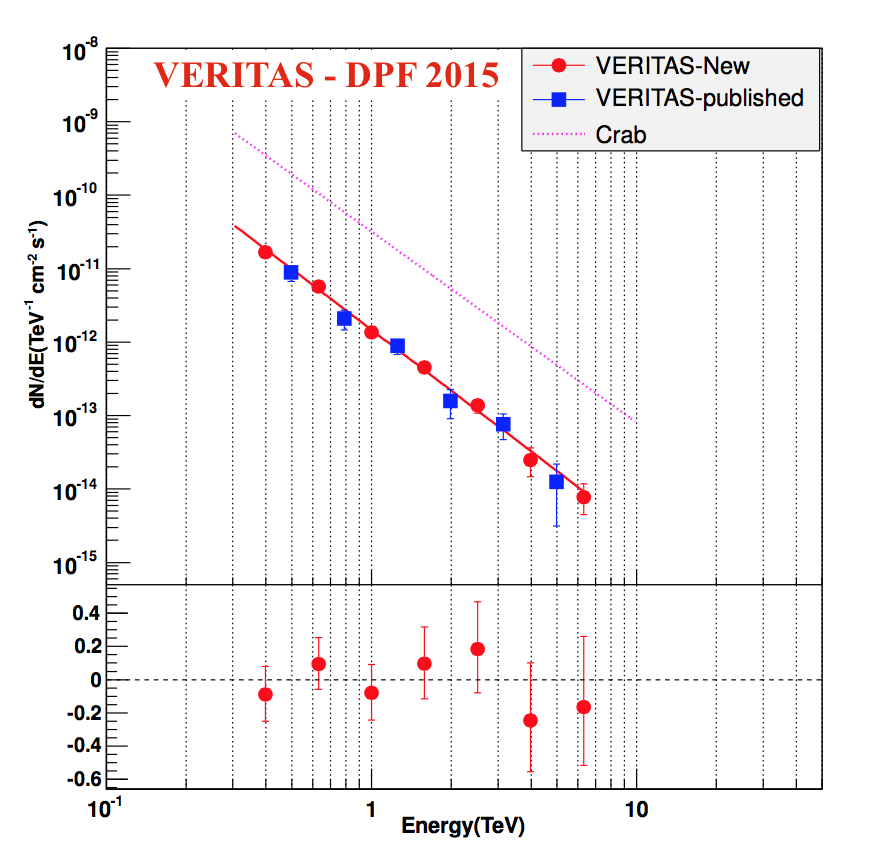}
\caption{Cas A differential energy spectrum: published VERITAS data~\cite{acciari} and new data fit to a power-law. The Crab spectrum is also shown for comparison.}
\label{fig:spectrum}
\end{figure}
%%%%%%%%%%%%%%%%%%%%%%%%%%%%%%%%%%%%%%%%%%%%%%%%%%%%%%%%%%%%%%%%%%%%%%%%%%%

Both the index and the normalization are in agreement with previous results by HEGRA~\cite{aharonian}, MAGIC~\cite{albert} and VERITAS~\cite{acciari}. 

\begin{figure}[htb]
\centering
\includegraphics[height=3in]{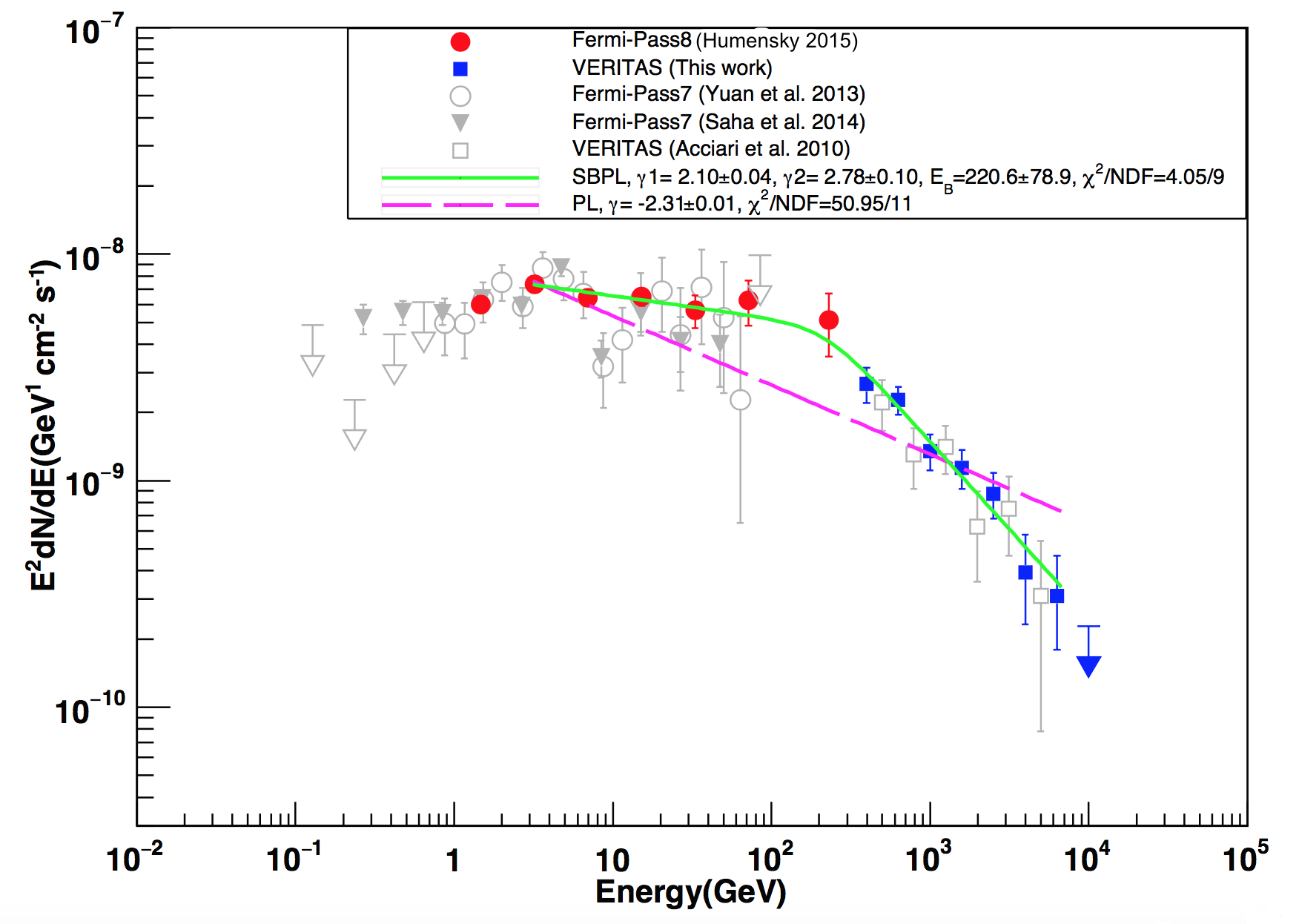}
\caption{for Cas A: Combined spectral points: Fermi-LAT and VERITAS current results. VERITAS points shown are for the entire data set. An upper limit is set at the 10 TeV bin~\cite{humensky} .}
\label{fig:fits}
\end{figure}

\begin{figure}[htb]
\centering
\includegraphics[height=3in]{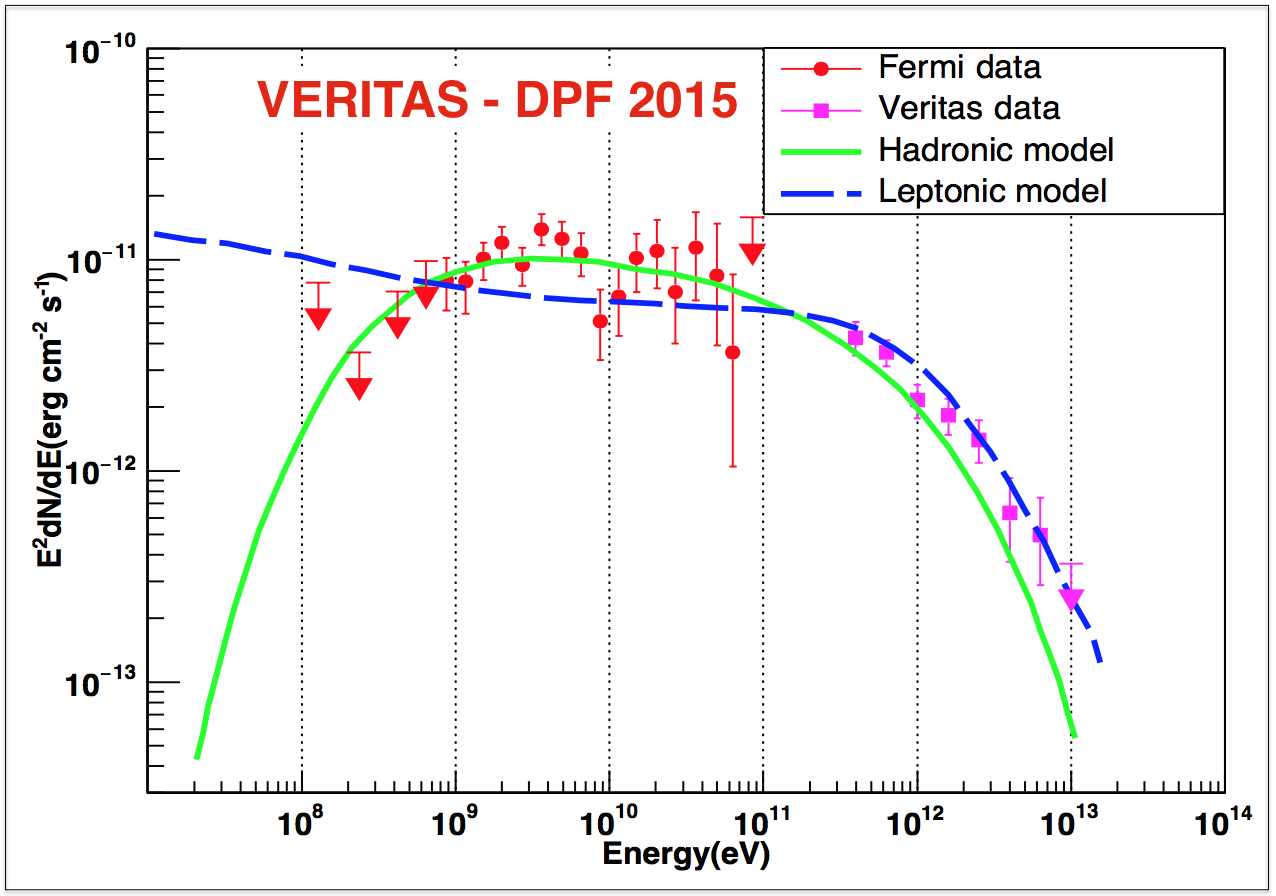}
\caption{Combined Fermi-LAT~\cite{yuan} and new VERITAS spectrum in comparison with leptonic and hadronic emission models ~\cite{abdo}.}
\label{fig:models}
\end{figure}

Figure \ref{fig:fits} shows the combined spectrum with Fermi-LAT~\cite{yuan, saha, humensky} and the complete VERITAS data set. Fitting the broad-band spectrum above 2 GeV, a broken power-law model is favored at the 4.9 sigma level over a single power-law when only statistical errors are considered, and at $>3.5$ sigma level when considering systematic errors as well. This suggests a softening of the spectrum above a few hundred GeV. 
In order for the data to be consistent with the current hadronic emission model (Figure \ref{fig:models}), the cut off energy would have to be higher than 10 TeV.

%%%%%%%%%%%%%%%%%%%%%%%%%%%%%%%%%%%%%%%%%%%%%%%%%%%%%%%%%%%%%%%%%%%%%%%%%
%%
%%   use this format to include an .pdf figure into your paper
%%%
%%%%%%%%%%%%%%%%%%%%%%%%%%%%%%%%%%%%%%%%%%%%%%%%%%%%%%%%%%%%%%%%%%%%%%%%%%%
 
%%%%%%%%%%%%%%%%%%%%%%%%%%%%%%%%%%%%%%%%%%%%%%%%%%%%%%%%%%%%%%%%%%%%%%%%%
%%
%%   use this format to include an .pdf figure into your paper
%%%
%%%%%%%%%%%%%%%%%%%%%%%%%%%%%%%%%%%%%%%%%%%%%%%%%%%%%%%%%%%%%%%%%%%%%%%%%%%

\begin{figure}[h]
\centering
\includegraphics[height=2.5in]{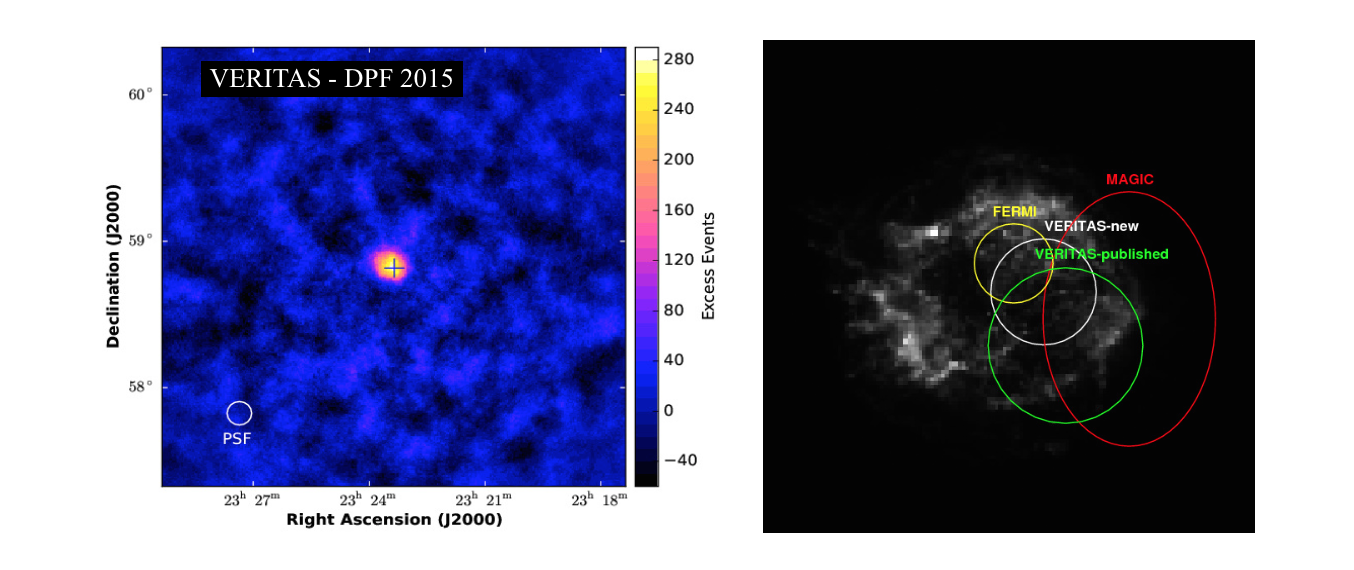}
\caption{Left: skymap with Cas A SZA data only, yielding a significance of $11 \sigma $. Point spread function (PSF) is shown as a white circle and the centroid location as a blue cross. Right: Chandra image of Cas A ~\cite{hwang} with overlayed comparison of centroid positions from Fermi (yellow, ~\cite{yuan}), VERITAS (green, ~\cite{acciari}) and MAGIC (red, ~\cite{albert}) with the new VERITAS (white). }
\label{fig:centroid}
\end{figure}

	Figure \ref{fig:centroid}  (left frame) shows a skymap generated by a reflected-region background model~\cite{berge}. It is derived from 18 h of post-upgrade observations at small zenith angles, yielding a significance of $11 \sigma $. Cas A is consistent with a point source for our point spread function (PSF) shown in white. The centroid is marked by a blue cross at $ RA=23\textrm{h}\
 23\textrm{m}\ 20.4\textrm{s}\ \pm \ 0\degree .006_{stat} \pm 0\degree .014_{sys} $ and $ Dec= 58.817 \pm 0\degree .006_{stat} \pm 0\degree .014_{sys} $. Figure \ref{fig:centroid} (right) compares the centroid positions from Fermi (yellow, ~\cite{yuan}), VERITAS (green, ~\cite{acciari}) and MAGIC (red, ~\cite{albert}) with the new VERITAS (white). This updated result is consistent with both the Fermi-LAT centroid location as well as our previously published centroid. This updated result does not confirm the speculation in ~\cite{grefenstette} that the GeV emission is associated with a north-central region bright in the infrared while the TeV emission is associated with bright synchrotron knots on the western side of the SNR.

%%%%%%%%%%%%%%%%%%%%%%%%%%%%%%%%%%%%%%%%%%%%%%%%%%%%%%%%%%%%%%%%%%%%%%%%%
%%
%%   use this format to include an .pdf figure into your paper
%%%
%%%%%%%%%%%%%%%%%%%%%%%%%%%%%%%%%%%%%%%%%%%%%%%%%%%%%%%%%%%%%%%%%%%%%%%%%%%

\section{Conclusions and Prospects}
We were able to refine the VHE spectrum, both at lower and higher energy.  Statistical errors in the index and in the centroid location are now below the current systematic uncertainties, motivating further work to improve our systematics. There are prospects for a better analysis process for the Large Zenith Angle data, important for the TeV-range spectrum.
Finally, in the near future the Cherenkov Telescope Array (CTA) will offer better angular resolution and pointing, which may refine the location of the centroid or even resolve the emission. CTA should also be able to extend the spectrum both to lower energy to overlap with Fermi-LAT and to energies beyond 10 TeV.

\Acknowledgments
   This research is supported by grants from the U.S. Department of Energy Office of Science, the U.S. National Science Foundation and the Smithsonian Institution, and by NSERC in Canada. We acknowledge the excellent work of the technical support staff at the Fred Lawrence Whipple Observatory and at the collaborating institutions in the construction and operation of the instrument. The VERITAS Collaboration is grateful to Trevor Weekes for his seminal contributions and leadership in the field of VHE gamma-ray astrophysics, which made this study possible.

\end{document}